\documentclass[useAMS,usenatbib]{mn2e}
\usepackage{times,amssymb}
\usepackage{graphicx,fleqn,colordvi}
\newcommand{\apj}{ApJ}                 
\newcommand{\mnras}{MNRAS}             \newcommand{\aap}{A\&A}

\newcommand{\prl}{Phys Rev Letters}    \newcommand{\pre}{Phys Rev E}
               \newcommand{\physrep}{Physics Reports}
\newcommand{\dd}{{\rm d}}
\newcommand{\vc}[1]{\textbf{\emph #1}}
\begin{document}
\label{firstpage}
\title[Solutions of statistical mechanics]{Second-order solutions of the equilibrium statistical mechanics for self-gravitating systems}
\author[P. He]{Ping He$$\thanks{E-mail: hep@itp.ac.cn}\\
Institute of Theoretical Physics, Chinese Academy of Sciences, Beijing 100190, China}
\date{\today}
\maketitle
\begin{abstract}
In a previous study, we formulated a framework of the entropy-based equilibrium statistical mechanics for self-gravitating systems. This theory is based on the Boltzmann-Gibbs entropy and includes the generalized virial equations as additional constraints. With the truncated distribution function to the lowest order, we derived a set of second-order equations for the equilibrium states of the system. In this work, the numerical solutions of these equations are investigated. It is found that there are three types of solutions for these equations. Both the isothermal and divergent solutions are thermally unstable and have unconfined density profiles with infinite mass, energy and spatial extent. The convergent solutions, however, seem to be reasonable. Although the results cannot match the simulation data well, because of the truncations of the distribution function and its moment equations, these lowest-order convergent solutions show that the density profiles of the system are confined, the velocity dispersions are variable functions of the radius, and the velocity distributions are also anisotropic in different directions. The convergent solutions also indicate that the statistical equilibrium of self-gravitating systems is by no means the thermodynamic equilibrium. These solutions are just the lowest-order approximation, but they have already manifested the qualitative success of our theory. We expect that higher-order solutions of our statistical-mechanical theory will give much better agreement with the simulation results concerning dark matter haloes.
\end{abstract}

\section{Introduction}
\label{sec:intro}

This is a follow-up work to our series of investigations of the statistical mechanics of self-gravitating systems. It has long been realized that the conventional methods of statistical mechanics of short-range interaction systems cannot be directly applied to the study long-range self-gravitating systems. Hence, it is necessary to return to the fundamentals of statistical mechanics and develop special techniques to handle the long-range nature of gravity \citep{padmanabhan90, chavanis06}. In our earlier two works \citep{hep10, kang11}, we performed preliminary investigations into the statistical mechanics of collisionless self-gravitating systems, and found that both the concept of entropy and the entropy principle are still valid for self-gravitating systems, which implies that an entropy-based statistical mechanics of self-gravitating systems is feasible.

In \citet{hep11}, we also investigated the stability of the statistical equilibrium of spherically symmetric collisionless self-gravitating systems. By calculating the second variation of the entropy, we found that perturbations of the relevant physical quantities should be classified as long- and short-range perturbations, which correspond to the long- and short-range relaxation mechanisms, respectively. We showed that the statistical equilibrium state with the most probable distribution of self-gravitating systems is neither maximum nor minimum, but is complex saddle-point entropy state, and hence differs greatly from the case of an ideal gas. Violent relaxation should be divided into two phases \citep[see][]{soker96}. The first phase is the entropy-production phase, while the second phase is the entropy-decreasing phase. The saddle-point entropy state is consistent with both \citet{ant62}'s proof and Binney's argument \citep[see][]{galdyn08}.
\begin{figure*}
\centerline{\includegraphics[width=1.75\columnwidth]{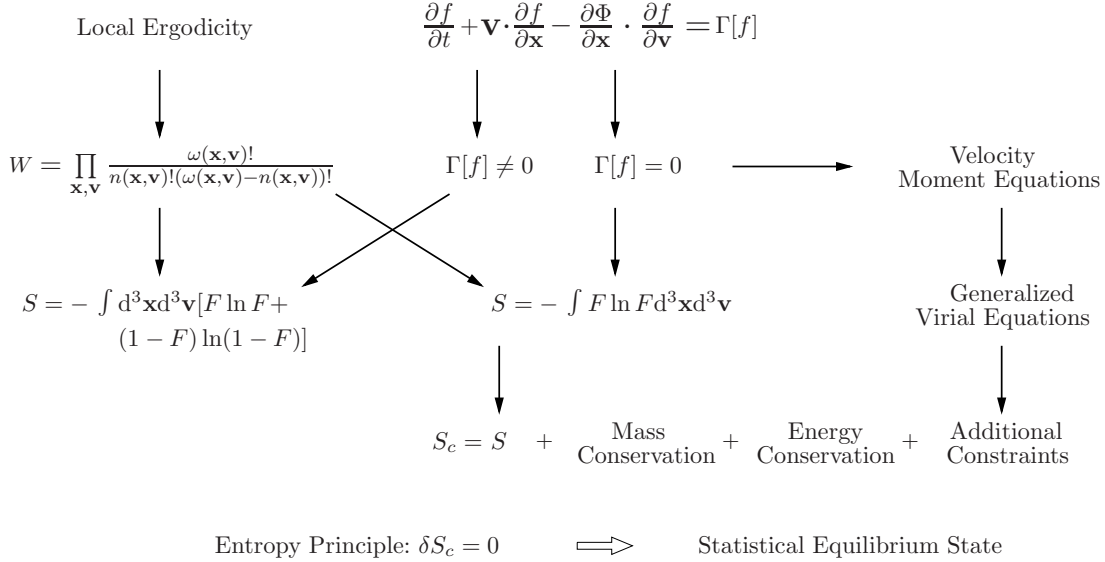}}
\caption{The framework of the equilibrium statistical mechanics for self-gravitating systems.}
\label{fig:framework}
\end{figure*}

Based on these findings, \citet{hep12} formulated a systematic theoretical framework of the statistical mechanics for spherically symmetric collisionless self-gravitating systems. It was demonstrated that the ergodicity is invalid globally for the whole self-gravitating system but can be re-established locally, and the Boltzmann-Gibbs entropy was re-derived, which should be the correct entropy form for collisionless gravitating systems. This framework is significantly different from previous works \cite[e.g.][]{lb67, shu78, kull97, tsallis88, nakamura00, hjorth10}, in that the collisionless Boltzmann equation (CBE) is included as an additional constraint on the entropy functional.

In this work, the aim is to study the lowest-order solutions of the statistical-mechanical theory we developed in \cite{hep12}. The paper is organized as follows. In Section~\ref{sec:statmech}, the framework of the statistical mechanical is briefly reviewed. Section~\ref{sec:results} presents the solutions of the second-order equations for the equilibrium states. A summary and the conclusions of our study are given in Section~\ref{sec:summary}.

\section{Framework of the statistical mechanics and its second-order equations of the statistical equilibrium state}
\label{sec:statmech}

\subsection{Framework of the statistical mechanics}
\label{ss:frame}

We outline the framework of the statistical mechanics of self-gravitating systems in Fig.~\ref{fig:framework}. The details of the theory can be found in \cite{hep12}.

\subsubsection{Local ergodicity and the Boltzmann-Gibbs entropy}
\label{sss:entropy}

Ergodicity, or the equiprobability principle, is fundamental for equilibrium statistical mechanics. However, we demonstrated that the ergodic hypothesis is not valid for the whole self-gravitating system. It seems that ergodicity breaking may be a common feature for long-range interaction systems \citep*[e.g.][]{mukamel05, campa09}. Fortunately, we can re-establish the local ergodicity in any local finitely small volume element of the system by treating gravitating particles as indistinguishable. With the validity of local ergodicity, we can still define entropy with the Boltzmann relation $S\equiv\ln W$, with $W$ being the Fermi-Dirac-like microstate number as
\begin{equation}
\label{eq:w}
W=\prod\limits_{\vc{x}, \vc{v}}\frac{\omega(\vc{x}, \vc{v})!} {n(\vc{ x}, \vc{v})! (\omega(\vc{x}, \vc{v}) - n(\vc{x}, \vc{v}))!},
\end{equation}
where $n (\vc{x}, \textbf {\emph v})$ is the velocity distribution in the volume element, and $\omega(\vc{x}, \vc{v})=\Delta^3\vc{x} \Delta^3 \vc{v}/h^3_g$ denotes the degeneracy at the velocity level $\vc{v}$ for the macrocell, with $h_g$ being the volume of the minimum phase-space element (i.e. microcell). As a result, the entropy-based equilibrium statistical mechanics for self-gravitating systems can be established.

The kinetic-theoretical equation for self-gravitating systems is the Boltzmann equation:
\begin{equation}
\label{eq:be}
\frac{\dd f}{\dd t} = \frac{\partial f}{\partial t} + \vc{v} \cdot \frac{\partial f}{\partial \vc{x}} - \frac{\partial \Phi}{\partial \vc{x}} \cdot \frac{\partial f}{\partial \vc{v}} = \Gamma[f].
\end{equation}
Here, $f(t, \vc{x}, \vc{v})$ is the {\em fine-grained} distribution function (DF), and $\Gamma[f]$ is the encounter operator, given by
\begin{equation}
\label{eq:gamma}
\Gamma [f({\bf w}_1, t)] \equiv N \int{\rm d}^6{\bf w}_2 \frac{\partial\Phi_{12}}
{\partial {\bf x}_1} \cdot \frac{\partial g({\bf w}_1, {\bf w}_2, t)}{\partial {\bf v}_1},
\end{equation}
where $\vc{w} = (\vc{x}, \vc{v})$ denotes the phase-space coordinates, $g$ is the two-body correlation function, and $\Phi_{12}$ is the gravitational potential between the two bodies $1$ and $2$ \cite[see equation~7.47 of][]{galdyn08}.

If either $\partial\Phi_{12}/\partial \vc{x}_1$ or $g$ is vanishing, so that $\Gamma[f]=0$, then the non-degenerate condition $n(\vc{x}, \vc{v})/\omega(\vc{x}, \vc{v}) \ll 1$ holds. With this non-degenerate condition, from equation~(\ref{eq:w}), we can derive the Boltzmann-Gibbs entropy:
\begin{equation}
\label{eq:bgs}
S[F] = -\int F(\vc{x}, \vc{v})\ln F(\vc{x}, \vc{v}) \dd^3\vc{x}\dd^3 \vc{v}.
\end{equation}
where $F (\vc{x}, \vc{v}) \equiv n(\vc{x},\vc{v})/\omega (\vc{x}, \vc{v})$ is the {\em coarse-grained} DF. However, if $\Gamma[f]\neq 0$, in this case the non-degenerate condition is invalid, and the correct entropy should be the Fermi-Dirac form as
\begin{displaymath}
S[F] = -\int \dd^3\vc{x}\dd^3\vc{v} [F(\vc{x}, \vc{v})\ln F(\vc{x}, \vc{v})
\end{displaymath}
\begin{equation}
\label{eq:fds}
{\hskip 10.25mm} + (1-F(\vc{x}, \vc{v}))\ln(1-F(\vc{x}, \vc{v}))].
\end{equation}

Because the CBE (i.e. equation~\ref{eq:be} with $\Gamma[f]=0$, also the Vlasov equation) is a good approximation for self-gravitating systems if the particle number $N$ of a system is very large, the Boltzmann-Gibbs entropy should be the proper entropy form for self-gravitating systems \citep*[cf.][]{tremaine86}.

The CBE might not be a good approximation in very dense regions, such as the center of a simulated $N$-body self-gravitating dark matter halo, where two-body relaxation might play an important role. In this case, the CBE would be better replaced by, say, the Balescu-Lenard equation \citep{heyvaerts10, chavanis10}, and the Fermi-Dirac-type entropy of equation~(\ref{eq:fds}), rather than the Boltzmann-Gibbs entropy, may be the correct entropy form.

\subsubsection{Additional macroscopic constraints}
\label{sss:addmac}

The statistical mechanics of self-gravitating systems also requires that the CBE should be considered in some way as the additional macroscopic constraint besides mass and energy conservation \citep{madsen87}, and \citet{hep12} has developed a way to properly manipulate this additional constraint for spherically symmetric self-gravitating systems. First, we integrate the steady-state CBE (i.e. $\partial f/\partial t = 0$ in equation~\ref{eq:be} with $\Gamma[f]=0$) to obtain its various orders of velocity-moment equations \citep[see equation~A11 of][]{hep12},
\begin{displaymath}
\frac{\dd}{\dd r}(\rho\overline{v^{k+2}_r v^m_{t}}) + \frac{(m+2)}{r} \rho\overline{v^{k+2}_r v^m_{t}} - \frac{(k+1)}{r}\frac{(m+2)}{(m+1)}
\end{displaymath}
\begin{equation}
\label{eq:meq}
{\hskip 5mm} \times \rho\overline{v^k_r v^{m+2}_{t}} = - (k+1)\frac{\dd \Phi}{\dd r} \rho\overline{v^k_r v^m_{t}},
\end{equation}
with the even integers $m,k\geq0$. These moment equations are then translated into their equivalent virialization forms \citep[see equation~9 of][]{hep12} as
\begin{displaymath}
- 2 (m^2-1)T_{k+2,m} + 2 (k+1)(m+2) T_{k,m+2}
\end{displaymath}
\begin{equation}
\label{eq:exvir}
{\hskip 2.5mm} + (k+1)(m+1)W_{k,m} = 4\pi (m+1) r^3 P_{k+2,m}.
\end{equation}
These generalized virial equations are just the proper forms for the additional macroscopic constraints besides mass and energy conservation, which is an extension to the original prescription by \citet{white87}.

Finally, we emphasis that the coarse-grained DF $F(\vc{x}, \vc{v})$ and the fine-grained DF $f(\vc{x}, \vc{v})$ are different, in that the entropy functional, equation~(\ref{eq:bgs}) or (\ref{eq:fds}), is constructed with $F$, but the DF in the CBE is $f$. There is no simple relationship between $F$ and $f$, and hence we should not directly include the steady-state CBE as the constraint in place of the series of generalized virial equations.

\begin{figure}
\centerline{\includegraphics[width=1.0\columnwidth]{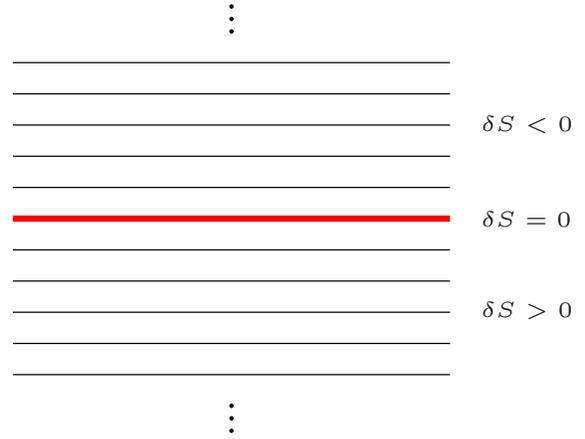}}
\caption{Illustration of existence of both the stationary states and the statistical equilibrium state. The thin lines indicate the stationary states, while the thick line indicates the statistical equilibrium state. The statistical equilibrium state is picked out by the entropy principle ($\delta S = 0$), which is exactly the saddle-point entropy state \citep{hep10, hep11}, whereas all the other stationary states do not satisfy the entropy principle ($\delta S \neq 0$).}
\label{fig:ss}
\end{figure}

\subsubsection{Out-of-equilibrium stationary states and the statistical equilibrium state}
\label{sss:qss}

As outlined above, the equilibrium states of self-gravitating systems consist of both mechanical and statistical equilibria, the former being characterized by a series of velocity-moment equations~(\ref{eq:meq}), and the latter by the statistical equilibrium equations, which should be derived from the entropy principle, $\delta S = 0$, with all the relevant macroscopic constraints being considered.

For a realistic gravitating system, however, such equilibrium states may not always be attained. Usually, if the collisionless relaxations, for example violent relaxation, do not last long enough, then the system will be trapped in so-called {\em quasi-stationary states} \citep[see][]{campa09, chavanis06}, whose lifetime depends on the total particle number $N$, proportional to $N/\ln N$. In the Vlasov limit, namely $N\rightarrow\infty$, this not-fully-relaxed system will finally be frozen into the out-of-equilibrium stable stationary state (i.e. $\delta S\neq 0$) and the statistical equilibrium state is never reached.

The statistical equilibrium state, characterized by the entropy principle $\delta S = 0$, is just a limit state, into which all the stationary states will converge, assuming that the collisionless relaxation lasts long enough. We depict the relationship between the stationary states and the statistical equilibrium state in Fig.~\ref{fig:ss}.

In reality, when a gravitating system, say a dark matter halo, experiences a monolithic collapse from the cold initial conditions \citep{white09} or repeated mergers of progenitor haloes \citep{white98}, it will, according to our viewpoint, attain the statistical equilibrium state (or at least reside in a state very close to this equilibrium state), whose mass-density profiles are described by well-known empirical universal laws \citep*{nfw97, moore99, einasto65}. If the systems are in quasi-stationary or stationary states, they may exhibit any possible density shapes, which differ from the universal laws \citep[e.g.][]{vanalbada82, jing00, zhang04, qin08, lemze11}.

Incidentally, the \citet{lb67} approach turns out to be a good way to explain qualitatively and quantitatively some features of quasi-stationary states \citep[see][and references therein]{campa09}. Statistical theories that quantitatively describe these quasi-stationary states have been constructed \citep[see][]{levin08a, levin08b, rizzato09}.

\subsection{Second-order equations of the equilibrium state}
\label{ss:lequ}

We can express the coarse-grained DF $F$ formally as the Taylor expansion
\begin{equation}
\label{eq:lmpd2}
F(\vc{x}, \vc{v}) = \exp\left(-\sum_{k,m,n} \lambda_{k,m,n} (\vc{x}) v^k_1 v^m_2 v^n_3\right),
\end{equation}
with
\begin{equation}
\label{eq:lagtlor}
\lambda_{k,m,n} (\vc{x}) = -\frac{1}{k!m!n!} \frac{\partial^{k+m+n} \ln F(\vc{ x}, \vc{v})} {\partial v^k_1 \partial v^m_2 \partial v^n_3}\big|_{\vc{v}=0},
\end{equation}
where $\lambda_{k,m,n} (\vc{x})$ are functions of the spatial coordinates $\vc{ x}$, and these undetermined coefficients should be determined by the joint mechanical and statistical equilibrium equations, and $k,m,n$ should be even integers, as there is no global momentum and angular momentum for the system.

We may expect that if all the expansion coefficients of $F(\vc{x}, \vc{v})$ in equation~(\ref{eq:lmpd2}) are exactly determined, the predictions will give accurate agreement with the simulation results. For realistic operations, however, we have to make some truncation to the DF.

In this work, we truncate $F(\vc{x}, \vc{v})$ of the spherically symmetric system to the second order (i.e. the lowest order) and we directly replace the expansion coefficients by the corresponding moments as
\begin{equation}
\label{eq:3df}
F(r, v_r, v_{\theta}, v_{\phi}) = \frac{\rho}{(2\pi)^{\frac{3}{2}}\sigma_r\sigma^2_{t}} {\rm exp} ( -\frac{v^2_r}{2 \sigma^2_r} - \frac{v^2_{\theta}}{2 \sigma^2_t} - \frac{v^2_{\phi}} {2 \sigma^2_t}).
\end{equation}
Here, the density $\rho$ and the velocity dispersions $\sigma^2_r$ and $\sigma^2_t$ are all functions of $r$. Because of the spherical symmetry, the dispersions of $v_{\theta}$ and $v_{\phi}$ are the same, so we uniformly use $\sigma^2_t$ to represent $\overline{v^2_{\theta}}$ or $\overline{v^2_{\phi}}$.

With this truncated DF, we use our theory of the statistical mechanics to derive the lowest-order (i.e. the second-order) equations as
\begin{equation}
\label{eq:je}
\frac{\dd p_r}{\dd r} + \frac{2}{r}(p_r - p_t) = - \rho \frac{G m}{r^2},
\end{equation}
\begin{equation}
\label{eq:ese1}
\frac{\dd\ln\rho}{\dd r}-\frac{\dd\ln p_t}{\dd r}-\frac{p_t}{r p_r}+\frac{1}{r}=0,
\end{equation}
\begin{equation}
\label{eq:ese2}
\frac{3}{2}\frac{\dd \ln\rho}{\dd r} - \frac{1}{p_t} \frac{\dd p_r}{\dd r}-\frac{2 p_r}{r p_t} + \frac{2}{r} = \lambda\frac{G m}{r^2},
\end{equation}
\begin{equation}
\label{eq:me}
\frac{\dd m}{\dd r} = 4\pi r^2 \rho,
\end{equation}
where $p_r=\rho\overline{v^2_r}$, and $p_t=\rho\overline{v^2_t}$ denotes both $p_{\theta}$ and $p_{\phi}$, with $\lambda$ in equation~(\ref{eq:ese2}) being the Lagrangian multiplier. Equation~(\ref{eq:je}) is just the Jeans equation \citep{galdyn08}, which is the lowest-order moment equation of equation~(\ref{eq:meq}) (i.e. $m=k=0$). Equations~(\ref{eq:ese1}) and (\ref{eq:ese2}) are two entropy stationary equations, derived from the entropy principle in \citet{hep12}. Equation~(\ref{eq:me}) is the differential form of the mass function, $m(r)=\int^r_0 4\pi r^2 \rho \dd r$. See section~4.3 of \citet{hep12} for details about these equations.

\section{Results}
\label{sec:results}

The purpose of developing the statistical mechanics of self-gravitating systems is to provide a physical account of all the empirical relationships found in cosmological simulations concerning dark matter haloes. The above equations, however, are just the lowest-order approximations, derived from the truncated DF of equation~(\ref{eq:lmpd2}). It should not be expected that these lowest-order equations will provide a good match to the simulation data. Nevertheless, these lowest approximations may still reveal some features of this statistical mechanics, and also be valuable in contributing to an understanding of the structure of self-gravitating dark haloes.

With a dimensional analysis, the Lagrangian multiplier $\lambda$ in equation~(\ref{eq:ese2}) can be written as
\begin{equation}
\label{eq:lambda}
\lambda = -\frac{\mu}{4\pi G \rho_s r^2_s},
\end{equation}
where $G$ is the Newtonian constant and $\mu$ is a dimensionless number. In all reasonable solutions, $\lambda$ should always be negative, and hence $\mu$ is positive. We can always set $\mu = 1$ by choosing appropriate values for the characteristic density $\rho_s$ and scale $r_s$. The initial radius is chosen as $r_i=10^{-7}r_s$. Given the initial values, we numerically solve these equations with the fourth-order Runge-Kutta method.

According to the initial conditions, we can classify the solutions into the following three types.

\begin{figure}
\centerline{\includegraphics[width=1.0\columnwidth]{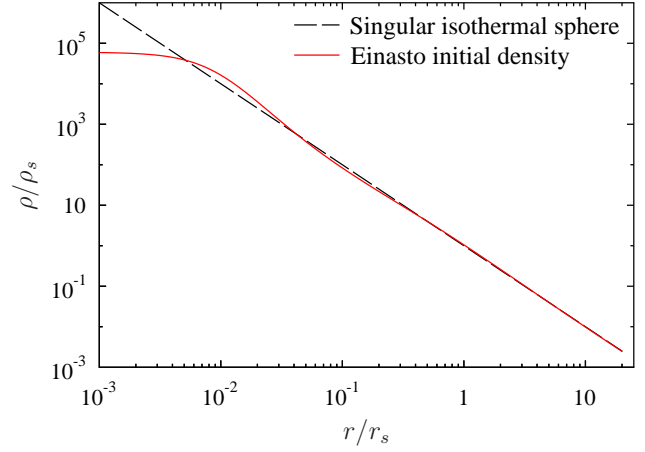}}
\caption{Density profiles of the isothermal solutions of equations~(\ref{eq:je}) -- (\ref{eq:me}). The behavior of the solutions depends on the initial values of $\rho_i$ in the inner regions, but all the solutions converge to the singular isothermal sphere $\rho \propto r^{-2}$ in the outskirts of the system.}
\label{fig:isothermal}
\end{figure}

\subsection{Isothermal solutions}
\label{ss:iso}

We define an effective velocity dispersion $\sigma^2_0$ as
\begin{equation}
\label{eq:sigma0}
\sigma^2_0\equiv 2\pi G \rho_s r^2_s.
\end{equation}
So if $\sigma^2_r=\sigma^2_t=\sigma^2_0$, the solutions of equations~(\ref{eq:je}) -- (\ref{eq:me}) are isothermal solutions, but the solutions of density $\rho(r)$ also depend on its initial values. For instance, if the initial value of $\rho_i$ at the radius $r_i$ is set as $\sigma^2_0/2\pi G r^2_i$, the equations will give the solution of the singular isothermal sphere $\rho \propto r^{-2}$, while if $\rho_i$ is set by the \citet{einasto65} profile
\begin{equation}
\label{eq:einasto}
\ln(\rho(r)/\rho_s)=-\frac{2}{\alpha}[(r/r_s)^{\alpha}-1],
\end{equation}
then the solution is flattened in the inner region and then converges to a singular isothermal sphere in the outskirts. We present the above two solutions in Fig.~\ref{fig:isothermal}.

Such a solution, similar to that of \citet{lb67}, has infinite mass, energy and spatial extent, and hence is not realistic. Moreover, it is well known that such an isothermal solution is thermally unstable, which will lead to gravothermal collapse (or gravothermal catastrophe) \citep{ant62, lb68}.

\begin{figure*}
\centerline{\includegraphics[width=1.75\columnwidth]{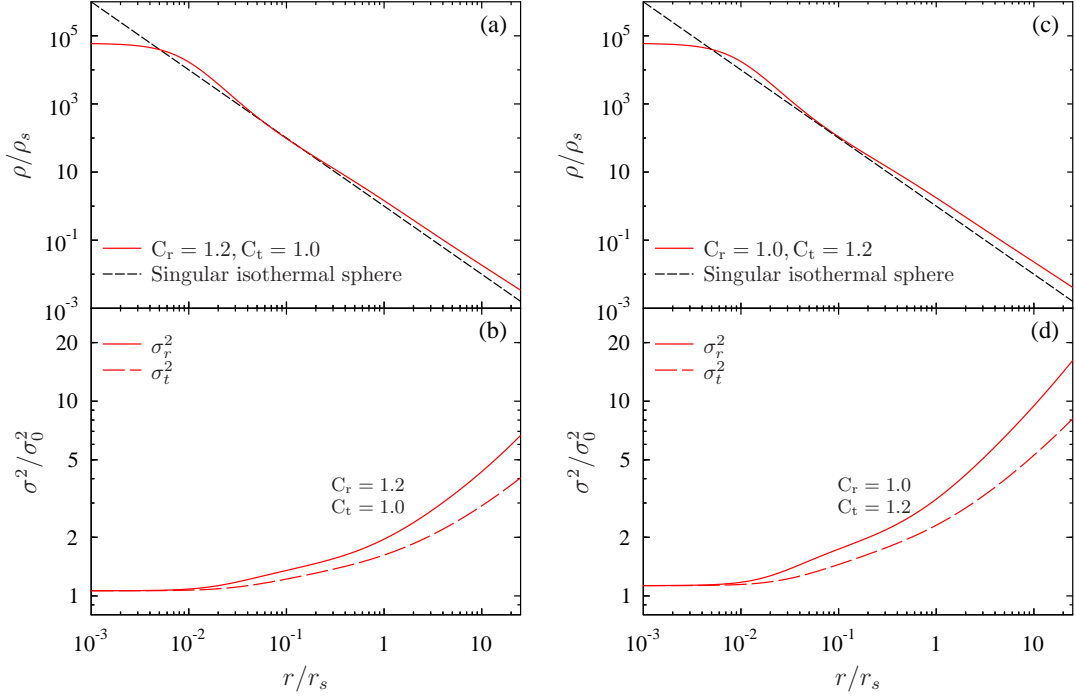}}
\caption{The divergent solutions of equations~(\ref{eq:je}) -- (\ref{eq:me}). The \citet{einasto65} density profile is used to set the initial value of $\rho_i$. The left two panels are for the case of ${\rm C_r} = 1.2$ and ${\rm C_t} =1$, and the right panels are for ${\rm C_r} = 1$ and ${\rm C_t} =1.2$. The dashed lines in panels (a) and (c) indicate the singular isothermal sphere, $\rho\propto r^{-2}$.}
\label{fig:upiso}
\end{figure*}

\begin{figure*}
\centerline{\includegraphics[width=1.75\columnwidth]{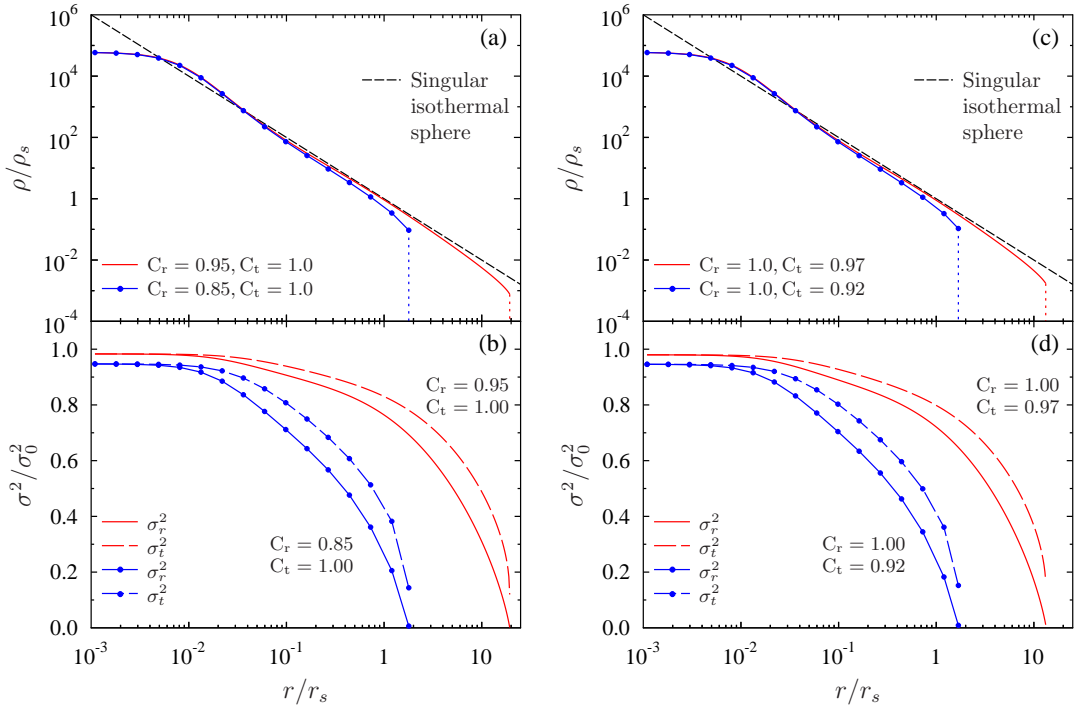}}
\caption{The convergent solutions of equations~(\ref{eq:je}) -- (\ref{eq:me}). The \citet{einasto65} density profile is used to set the initial value of $\rho_i$. The left two panels are for the case of ${\rm C_r} < 1$, and the right panels are for ${\rm C_t} <1$. The dotted lines at the ends of the density functions indicate the truncation caused by the vanishing radial velocity dispersion $\sigma^2_r$. Different initial values of ${\rm C_r}$ and ${\rm C_t}$ are indicated in the figure. The dashed lines in panels (a) and (c) indicate the singular isothermal sphere, $\rho\propto r^{-2}$.}
\label{fig:lowiso}
\end{figure*}

\subsection{Divergent solutions}
\label{ss:div}

We define the ratios of the radial and tangential velocity dispersion over the effective velocity dispersion $\sigma^2_0$ as
\begin{equation}
\label{eq:vdratio}
{\rm C_r} = \frac{\sigma^2_r}{\sigma^2_0}, {\hskip 10mm} {\rm C_t} = \frac{\sigma^2_t}{\sigma^2_0}.
\end{equation}
So for the isothermal solution, ${\rm C_r} = {\rm C_t} = 1$. Now we treat the initial ${\rm C_r}$ and ${\rm C_t}$ at $r_i$ as two free parameters, and see how the different initial values of ${\rm C_r}$ and ${\rm C_t}$ will affect the behavior of the solutions. In Fig.~\ref{fig:upiso}, we present two solutions with either ${\rm C_r}$ or ${\rm C_t}$
being larger than 1. We can see that the density functions of both the solutions go outwards with respect to the radius $r$ at a shallower slope than that of the singular isothermal sphere, and that all the velocity dispersions increase with $r$. So these solutions are divergent and unconfined, and of course are even more thermally unstable than the isothermal solutions.

\subsection{Convergent solutions}
\label{ss:con}

If either ${\rm C_r}$ or ${\rm C_t}$ is smaller than 1, the solutions will be convergent and finite. In Fig.~\ref{fig:lowiso}, we show these solutions with different initial values of ${\rm C_r}$ and ${\rm C_t}$. The density functions at the outer boundary of the system are automatically truncated as a result of the vanishing radial velocity dispersions. Although the results cannot match the simulations well \citep[e.g.][]{navarro10, ludlow10}, these lowest-order solutions do possess the following properties: (i) the density profile is finite, which does not lead to a state with infinite mass or energy; (ii) the velocity dispersions are variable functions of the radius $r$, and (iii) the velocity distributions are anisotropic in different directions. The last two points indicate that the statistical equilibrium of self-gravitating systems is by no means the thermodynamic equilibrium.

We use the Boltzmann-Gibbs entropy in our statistical-mechanical theory. It might be a misunderstanding that the Boltzmann-Gibbs statistics must lead to the Maxwell-Boltzmann velocity distribution. However, by including the additional constraints of the generalized virial equations~(\ref{eq:exvir}), the Boltzmann-Gibbs statistics is endowed with new properties, in that the theory may give any possible velocity distributions.

As addressed previously, these convergent solutions are just the lowest-order approximations, derived from the truncations of the DF, equation~(\ref{eq:lmpd2}), and its moment equations~(\ref{eq:meq}). As a result, these lowest-order equations cannot provide a good match to the simulation data. We are optimistic in expecting that higher-order solutions of the statistical-mechanical theory will give much better agreement with the realistic data.

\section{Summary and Conclusions}
\label{sec:summary}

In \citet{hep12}, a systematic theoretical framework was formulated for the statistical mechanics of spherically symmetric collisionless self-gravitating systems based on the Boltzmann-Gibbs entropy. According to the statistical-mechanical theory, the equilibrium states of self-gravitating systems consist of both mechanical and statistical equilibria. The mechanical equilibria are characterized by the steady-state Vlasov equation, or its equivalent velocity-moment equations, and the statistical equilibrium equations should be derived from the entropy principle.

For a realistic gravitating system, however, such equilibrium states may not always be attained. Usually, if the collisionless relaxations do not last long enough, the system will be trapped in quasi-stationary states, whose lifetimes depend on the total particle number. In the Vlasov limit, this not-fully-relaxed system will finally be frozen into the out-of-equilibrium stable stationary state, and the statistical equilibrium state is never reached. Hence, the statistical equilibrium state is just a limit state, to which all the stationary states will converge, if that the collisionless relaxation lasts long enough.

The coarse-grained DF is formally expressed as a Taylor series, and if all the expansion coefficients are determined, the theory will give exact predictions. For practical purposes, however, we have to truncate this Taylor series. In this work, we truncate the DF for the spherically symmetric system to the second order (i.e. the lowest order). With this truncated DF, we use our theory to derive a set of second-order equations, and it was our aim to solve these equations in this paper.

According to the initial conditions, we can classify the solutions into isothermal, divergent and convergent solutions. We find that the isothermal solutions, similar to that of \citet{lb67}, have infinite mass, energy and spatial extent, and hence are not acceptable. Moreover, it is well known that such isothermal solutions are thermally unstable, which will lead to gravothermal collapse for the system. The divergent solutions should be even more thermally unstable than the isothermal solution.

The convergent solutions seem, however, to be reasonable. Although the results cannot match the simulations well, these lowest-order solutions show that the density profiles are finite and confined, the velocity dispersions are variable functions of $r$, and the velocity distributions are also anisotropic. In addition, the solutions indicate that the statistical equilibrium of self-gravitating systems is by no means the thermodynamic equilibrium.

Our statistical-mechanical theory is based on the Boltzmann-Gibbs entropy. By including the generalized virial equations as additional macroscopic constraints, this new Boltzmann-Gibbs statistics may give any possible velocity distributions. These convergent solutions are just the lowest-order approximations, derived from the truncations of the DF and its moment equations, but they have already manifested the qualitative success of our theory. We expect that higher-order solutions of the statistical-mechanical theory will give much better agreement with the simulation results for dark matter haloes.

\section*{Acknowledgements}
The author would like to thank the referee and the editor for their helpful suggestions and comments. This work is supported by National Basic Research Program of China, No:2010CB832805.


\label{lastpage}
\end{document}